
\input phyzzx
%
\newbox\hdbox%
\newcount\hdrows%
\newcount\multispancount%
\newcount\ncase%
\newcount\ncols
\newcount\nrows%
\newcount\nspan%
\newcount\ntemp%
\newdimen\hdsize%
\newdimen\newhdsize%
\newdimen\parasize%
\newdimen\spreadwidth%
\newdimen\thicksize%
\newdimen\thinsize%
\newdimen\tablewidth%
\newif\ifcentertables%
\newif\ifendsize%
\newif\iffirstrow%
\newif\iftableinfo%
\newtoks\dbt%
\newtoks\hdtks%
\newtoks\savetks%
\newtoks\tableLETtokens%
\newtoks\tabletokens%
\newtoks\widthspec%
%
%
\immediate\write15{%
CP SMSG GJMSINK TEXTABLE --> TABLE MACROS V. 851121 JOB = \jobname%
}%
%
%
\tableinfotrue%
\catcode`\@=11
%
%
\def\tstrut{\vrule height3.1ex depth1.2ex width0pt}%
\def\and{\char`\&}
\def\tablerule{\noalign{\hrule height\thinsize depth0pt}}%
\thicksize=1.5pt
\thinsize=0.6pt
\def\thickrule{\noalign{\hrule height\thicksize depth0pt}}%
\def\ctr#1{\hfil\ #1\hfil}%
%
%
%
%
\tablewidth=-\maxdimen%
\spreadwidth=-\maxdimen%
\def\tabskipglue{0pt plus 1fil minus 1fil}%
%
%
\centertablestrue%
%
%
%
%
\parasize=4in%
\gdef\ARGS{########}
\gdef\headerARGS{####}
\def\@mpersand{&}
{\catcode`\|=13
\gdef\letbarzero{\let|0}
\gdef\letbartab{\def|{&&}}%
\gdef\letvbbar{\let\vb|}%
}
{\catcode`\&=4
\def\ampskip{&\omit\hfil&}
\catcode`\&=13
\let&0
\xdef\letampskip{\def&{\ampskip}}%
\gdef\letnovbamp{\let\novb&\let\tab&}
}
\def\begintable{
   \begingroup%
   \catcode`\|=13\letbartab\letvbbar%
   \catcode`\&=13\letampskip\letnovbamp%
   \def\multispan##1{
      \omit \mscount##1%
      \multiply\mscount\tw@\advance\mscount\m@ne%
      \loop\ifnum\mscount>\@ne \sp@n\repeat%
   }
   \def\|{%
      &\omit\widevline&%
   }%
   \ruledtable
}
\long\def\ruledtable#1\endtable{%
%
%
%
   \offinterlineskip
   \tabskip 0pt
   \def\widevline{\vrule width\thicksize}
   \def\endrow{\@mpersand\omit\hfil\crnorm\@mpersand}%
   \def\crthick{\@mpersand\crnorm\thickrule\@mpersand}%
   \def\crthickneg##1{\@mpersand\crnorm\thickrule
          \noalign{{\skip0=##1\vskip-\skip0}}\@mpersand}%
   \def\crnorule{\@mpersand\crnorm\@mpersand}%
   \def\crnoruleneg##1{\@mpersand\crnorm
          \noalign{{\skip0=##1\vskip-\skip0}}\@mpersand}%
   \let\nr=\crnorule
   \def\endtable{\@mpersand\crnorm\thickrule}%
   \let\crnorm=\cr
%
%
   \edef\cr{\@mpersand\crnorm\tablerule\@mpersand}%
   \def\crneg##1{\@mpersand\crnorm\tablerule
          \noalign{{\skip0=##1\vskip-\skip0}}\@mpersand}%
   \let\ctneg=\crthickneg
   \let\nrneg=\crnoruleneg
   \the\tableLETtokens
%
%
   \tabletokens={&#1}
%
%
   \countROWS\tabletokens\into\nrows%
   \countCOLS\tabletokens\into\ncols%
%
%
   \advance\ncols by -1%
   \divide\ncols by 2%
   \advance\nrows by 1%
%
%
   \iftableinfo %
      \immediate\write16{[Nrows=\the\nrows, Ncols=\the\ncols]}%
   \fi%
%
%
   \ifcentertables
      \ifhmode \par\fi
      \hbox to \hsize{
      \hss
   \else %
      \hbox{%
   \fi
      \vbox{%
         \makePREAMBLE{\the\ncols}
         \edef\next{\preamble}
         \let\preamble=\next
         \makeTABLE{\preamble}{\tabletokens}
      }
      \ifcentertables \hss}\else }\fi
   \endgroup
   \tablewidth=-\maxdimen
   \spreadwidth=-\maxdimen
}
\def\makeTABLE#1#2{
   {
   \let\ifmath0
   \let\header0
   \let\multispan0
%
%
   \ncase=0%
   \ifdim\tablewidth>-\maxdimen \ncase=1\fi%
   \ifdim\spreadwidth>-\maxdimen \ncase=2\fi%
   \relax
%
   \ifcase\ncase %
      \widthspec={}%
   \or %
      \widthspec=\expandafter{\expandafter t\expandafter o%
                 \the\tablewidth}%
   \else %
      \widthspec=\expandafter{\expandafter s\expandafter p\expandafter r%
                 \expandafter e\expandafter a\expandafter d%
                 \the\spreadwidth}%
   \fi %
   \xdef\next{
      \halign\the\widthspec{%
      #1
      \noalign{\hrule height\thicksize depth0pt}
      \the#2\endtable
%
      }
   }
   }
   \next
}
\def\makePREAMBLE#1{
   \ncols=#1
   \begingroup
   \let\ARGS=0
   \edef\xtp{\widevline\ARGS\tabskip\tabskipglue%
   &\ctr{\ARGS}\tstrut}
   \advance\ncols by -1
   \loop
      \ifnum\ncols>0 %
      \advance\ncols by -1%
      \edef\xtp{\xtp&\vrule width\thinsize\ARGS&\ctr{\ARGS}}%
   \repeat
   \xdef\preamble{\xtp&\widevline\ARGS\tabskip0pt%
   \crnorm}
   \endgroup
}
\def\countROWS#1\into#2{
   \let\countREGISTER=#2%
   \countREGISTER=0%
   \expandafter\ROWcount\the#1\endcount%
}%
\def\ROWcount{%
   \afterassignment\subROWcount\let\next= %
}%
\def\subROWcount{%
   \ifx\next\endcount %
      \let\next=\relax%
   \else%
      \ncase=0%
      \ifx\next\cr %
         \global\advance\countREGISTER by 1%
         \ncase=0%
      \fi%
      \ifx\next\endrow %
         \global\advance\countREGISTER by 1%
         \ncase=0%
      \fi%
      \ifx\next\crthick %
         \global\advance\countREGISTER by 1%
         \ncase=0%
      \fi%
      \ifx\next\crnorule %
         \global\advance\countREGISTER by 1%
         \ncase=0%
      \fi%
      \ifx\next\crthickneg %
         \global\advance\countREGISTER by 1%
         \ncase=0%
      \fi%
      \ifx\next\crnoruleneg %
         \global\advance\countREGISTER by 1%
         \ncase=0%
      \fi%
      \ifx\next\crneg %
         \global\advance\countREGISTER by 1%
         \ncase=0%
      \fi%
      \ifx\next\header %
         \ncase=1%
      \fi%
      \relax%
      \ifcase\ncase %
         \let\next\ROWcount%
      \or %
         \let\next\argROWskip%
      \else %
      \fi%
   \fi%
   \next%
}
\def\counthdROWS#1\into#2{%
\dvr{10}%
   \let\countREGISTER=#2%
   \countREGISTER=0%
\dvr{11}%
\dvr{13}%
   \expandafter\hdROWcount\the#1\endcount%
\dvr{12}%
}%
\def\hdROWcount{%
   \afterassignment\subhdROWcount\let\next= %
}%
\def\subhdROWcount{%
   \ifx\next\endcount %
      \let\next=\relax%
   \else%
      \ncase=0%
      \ifx\next\cr %
         \global\advance\countREGISTER by 1%
         \ncase=0%
      \fi%
      \ifx\next\endrow %
         \global\advance\countREGISTER by 1%
         \ncase=0%
      \fi%
      \ifx\next\crthick %
         \global\advance\countREGISTER by 1%
         \ncase=0%
      \fi%
      \ifx\next\crnorule %
         \global\advance\countREGISTER by 1%
         \ncase=0%
      \fi%
      \ifx\next\header %
         \ncase=1%
      \fi%
\relax%
      \ifcase\ncase %
         \let\next\hdROWcount%
      \or%
         \let\next\arghdROWskip%
      \else %
      \fi%
   \fi%
   \next%
}%
{\catcode`\|=13\letbartab
\gdef\countCOLS#1\into#2{%
   \let\countREGISTER=#2%
   \global\countREGISTER=0%
   \global\multispancount=0%
   \global\firstrowtrue
   \expandafter\COLcount\the#1\endcount%
   \global\advance\countREGISTER by 3%
   \global\advance\countREGISTER by -\multispancount
}%
\gdef\COLcount{%
   \afterassignment\subCOLcount\let\next= %
}%
{\catcode`\&=13%
\gdef\subCOLcount{%
   \ifx\next\endcount %
      \let\next=\relax%
   \else%
      \ncase=0%
      \iffirstrow
         \ifx\next& %
            \global\advance\countREGISTER by 2%
            \ncase=0%
         \fi%
         \ifx\next\span %
            \global\advance\countREGISTER by 1%
            \ncase=0%
         \fi%
         \ifx\next| %
            \global\advance\countREGISTER by 2%
            \ncase=0%
         \fi
         \ifx\next\|
            \global\advance\countREGISTER by 2%
            \ncase=0%
         \fi
         \ifx\next\multispan
            \ncase=1%
            \global\advance\multispancount by 1%
         \fi
         \ifx\next\header
            \ncase=2%
         \fi
         \ifx\next\cr       \global\firstrowfalse \fi
         \ifx\next\endrow   \global\firstrowfalse \fi
         \ifx\next\crthick  \global\firstrowfalse \fi
         \ifx\next\crnorule \global\firstrowfalse \fi
         \ifx\next\crnoruleneg \global\firstrowfalse \fi
         \ifx\next\crthickneg  \global\firstrowfalse \fi
         \ifx\next\crneg       \global\firstrowfalse \fi
      \fi
\relax
      \ifcase\ncase %
         \let\next\COLcount%
      \or %
         \let\next\spancount%
      \or %
         \let\next\argCOLskip%
      \else %
      \fi %
   \fi%
   \next%
}%
\gdef\argROWskip#1{%
   \let\next\ROWcount \next%
}
\gdef\arghdROWskip#1{%
   \let\next\ROWcount \next%
}
\gdef\argCOLskip#1{%
   \let\next\COLcount \next%
}
}
}
\def\spancount#1{
   \nspan=#1\multiply\nspan by 2\advance\nspan by -1%
   \global\advance \countREGISTER by \nspan
   \let\next\COLcount \next}%
\def\dvr#1{\relax}%
\def\header#1{%
\dvr{1}{\let\cr=\@mpersand%
\hdtks={#1}%
\counthdROWS\hdtks\into\hdrows%
\advance\hdrows by 1%
\ifnum\hdrows=0 \hdrows=1 \fi%
\dvr{5}\makehdPREAMBLE{\the\hdrows}%
\dvr{6}\getHDdimen{#1}%
{\parindent=0pt\hsize=\hdsize{\let\ifmath0%
\xdef\next{\valign{\headerpreamble #1\crnorm}}}\dvr{7}\next\dvr{8}%
}%
}\dvr{2}}
\def\makehdPREAMBLE#1{
\dvr{3}%
\hdrows=#1
{
\let\headerARGS=0%
\let\cr=\crnorm%
\edef\xtp{\vfil\hfil\hbox{\headerARGS}\hfil\vfil}%
\advance\hdrows by -1
\loop
\ifnum\hdrows>0%
\advance\hdrows by -1%
\edef\xtp{\xtp&\vfil\hfil\hbox{\headerARGS}\hfil\vfil}%
\repeat%
\xdef\headerpreamble{\xtp\crcr}%
}
\dvr{4}}
\def\getHDdimen#1{%
\hdsize=0pt%
\getsize#1\cr\end\cr%
}
\def\getsize#1\cr{%
\endsizefalse\savetks={#1}%
\expandafter\lookend\the\savetks\cr%
\relax \ifendsize \let\next\relax \else%
\setbox\hdbox=\hbox{#1}\newhdsize=1.0\wd\hdbox%
\ifdim\newhdsize>\hdsize \hdsize=\newhdsize \fi%
\let\next\getsize \fi%
\next%
}%
\def\lookend{\afterassignment\sublookend\let\looknext= }%
\def\sublookend{\relax%
\ifx\looknext\cr %
\let\looknext\relax \else %
   \relax
   \ifx\looknext\end \global\endsizetrue \fi%
   \let\looknext=\lookend%
    \fi \looknext%
}%
%
%
\def\tablelet#1{%
   \tableLETtokens=\expandafter{\the\tableLETtokens #1}%
}%
\catcode`\@=12

\def\({[}
\def\){]}
\def\e{{\epsilon}}
\def\ea{{\epsilon_1}}
\def\eb{{\epsilon_2}}
\def\NP{{\it Nucl. Phys.}~}
\def\PR{{\it Phys. Rev.}~}
\def\PRL{{\it Phys. Rev. Lett.}~}
\def\PL{{\it Phys. Lett.}~}
\def\PHRE{{\it Phys. Rep.}~}
\def\tM{{\tilde M}}
\def\tm{{\tilde m}}
\def\tV{{\tilde V}}

\REF\pdg{K. Hikasa {\it et al.}, Particle Data Group,  \PR {\bf D45}
(1992) S1.}
\REF\bsg{E. Thorndike, CLEO Collaboration, a talk given in
the meeting of the American Physical Society, Washington D.C. (1993).}
\REF\gama{F. Gabbiani and A. Masiero, \NP {\bf B322} (1989) 235.}
\REF\nilles{H.P. Nilles, \PHRE {\bf 110} (1984) 1.}
\REF\georgi{H. Georgi, \PL {\bf 169B} (1986) 231;
L.J. Hall, V.A. Kostelecky and S. Raby, \NP {\bf B267} (1986) 415.}
\REF\weinberg{L. Hall, J. Lykken and S. Weinberg, \PR {\bf D27} (1983)
2359.}
\REF\ibanez{M. Dine, A. Kagan and S. Samuel, \PL {\bf 243B} (1990) 250;
L. Ibanez and D. Lust, \NP {\bf B382} (1992) 305;
B. de Carlos, J.A. Casas and C. Munoz, \PL {\bf 299B} (1993) 234;
CERN-TH.6436/92;
V. Kaplunovski and J. Louis, CERN-TH.6809/93 (1993).}
\REF\dine{M. Dine and A.E. Nelson, SCIPP 93/03 (1993).}
\REF\dinetoo{M. Dine, A. Kagan and R. Leigh, SCIPP 93/04 (1993).}
\REF\frni{C.D. Froggatt and H.B. Nielsen, \NP {\bf B147} (1979) 277.}
\REF\dim{Z.G. Berezhiani, \PL {\bf 129B} (1983) 99; {\bf 150B} (1985)
177;
S. Dimopoulos, \PL {\bf 129B} (1983) 417;
J. Bagger, S. Dimopoulos, E. Masso and M. Reno,\NP {\bf B258} (1985)
565;
J. Bagger, S. Dimopoulos, H. Georgi and S. Raby, In: {\it Proc. Fifth
Workshop on Grand Unification.} Eds. Kang, K., Fried, H. and Frampton,
P., Singapore, World Scientific (1984);
A. Davidson, V.P. Nair and K.C. Wali, \PR {\bf D29} (1984) 1505;
A. Davidson and K.C. Wali, \PRL {\bf 60} (1988) 1313;
A. Davidson, S. Ranfone and K.C. Wali, \PR {\bf D41} (1990) 208.}
\REF\lns{M. Leurer, Y. Nir and N. Seiberg, RU-92/59, \NP {\bf B}, in
press.}
\REF\dght{J. F. Donoghue, E. Golowich, B. R. Holstein and J. Trampetic,
\PR {\bf D33} (1986) 179.}
\REF\georgid{H. Georgi, HUTP-92/A049 (1992);
T. Ohl, G. Ricciardi, E.H. Simmons, HUTP-92/A053.}

\def\Weizmann{\centerline{\it Weizmann Institute of Science}
\centerline{\it Physics Department, Rehovot 76100, Israel}}
\def\Rutgers{\centerline{\it Department of Physics and Astronomy}
\centerline{\it Rutgers University, Piscataway, NJ 08855-0849, USA}}
{\baselineskip=11pt
\Pubnum={RU-93-16 \cr WIS-93/37/Apr-PH}
\date={April, 1993}
\titlepage
\title{{\bf Should Squarks Be Degenerate?}}
\author{Yosef Nir}
\Weizmann
\centerline{ftnir@weizmann.bitnet}
\vskip .2in
\andauthor{Nathan Seiberg}
\Rutgers
\centerline{seiberg@physics.rutgers.edu}

\vskip .2in

\centerline{\bf Abstract}
For generic squark masses, box diagrams with squarks and gluinos give
unacceptably large contributions to neutral meson ($K$, $B$ and $D$)
mixing. The standard solution to this problem is to assume that squarks
are degenerate to a very good approximation. We suggest an alternative
mechanism to suppress squark contributions to flavor changing neutral
currents: the alignment of quark with squark mass matrices. This
mechanism arises naturally in the framework of Abelian horizontal
symmetries.
\vfill
\endpage
}

Within the Standard Model, flavor changing neutral current (FCNC)
interactions are highly suppressed by the weak coupling constant, by
small mixing angles and by small fermion masses. This makes such
processes a particularly sensitive probe of new physics at high energy
scales. New contributions to FCNC may be comparable to or even dominate
over the Standard Model contributions. Thus, measurements of FCNC
processes such as neutral meson mixing [\pdg],
$$\eqalign{ {\Delta m_K\over m_K}=&\ 7\times10^{-15},\cr {\Delta
m_B\over m_B}\approx&\ 7\times10^{-14},\cr {\Delta m_D\over m_D}\leq&\
7\times10^{-14},\cr} \eqn\mixing$$
and radiative $B$-decay [\bsg],
$$BR(B\rightarrow X_s\gamma)\leq5.4\times10^{-4},\eqn\radiative$$
put severe constraints on extensions of the Standard Model.

Supersymmetric extensions of the Standard Model predict large new
contributions to FCNC processes. Squarks and gluinos contribute to
\mixing\ through box diagrams and to \radiative\ through penguin
diagrams. A possible suppression due to large squark and gluino masses
is easily compensated for by three enhancement factors:

$(i)$ Matrix elements of new four quark operators are enhanced due to
different Lorentz structure;

$(ii)$ The weak coupling of the Standard Model diagrams is replaced by
the strong coupling;

$(iii)$ The GIM mechanism does not operate for generic squark masses.

The resulting contributions are so large, even for squark masses as
heavy as $1\ TeV$, that the processes \mixing\ and \radiative\ severely
constrain the form of the squark mass matrices.

A convenient way to present these constraints is the following.  We
denote the quark mass matrices and squark mass-squared matrices by $M^q$
and $\tilde M^{q2}$, respectively ($q=u,d$).  Supersymmetry correlates
the bases of the quarks and the squarks but in general there is no
preferred basis for these superfields.  However, as we will see below,
models with horizontal symmetry have such a preferred basis.  The $M^q$
matrices are diagonalized by bi-unitary transformations:
$$\eqalign{V_L^u M^u V_R^{u\dagger}=&{\rm diag}\{m_u,m_c,m_t\},\cr V_L^d
M^d V_R^{d\dagger}=&{\rm diag}\{m_d,m_s,m_b\}.\cr}\eqn\defv$$
The $6\times6$ matrices $\tilde M^{q2}$ can be divided into $3\times3$
sub-matrices:
$$\tilde M^{q2}=\pmatrix{\tilde M^{q2}_{LL} & \tilde M^{q2}_{LR} \cr
\tilde M^{q2\dagger}_{LR} & \tilde M^{q2}_{RR} \cr}\eqn\defsubm$$
where $\tilde M^{q2}_{LL}$ and $\tilde M^{q2}_{RR}$
are hermitian matrices.

If all off-diagonal terms in $\tilde M^{q2}$ are smaller than the
diagonal ones, and $V_{L,R}^q$ are close to the identity, then the
quantities constrained by FCNC processes are
$$(\delta^q_{MN})_{ij}={(V_M^q \tilde M^{q2}_{MN}
V_N^{q\dagger})_{ij}\over\tilde m^2}\eqn\defdelta$$
(here and below $M$ and $N$ run over $L$ and $R$).
The dimensionless $\delta^q_{MN}$-matrices have the simple meaning of
squark mass-squared matrices (normalized to the average squark
mass-squared $\tm^2$) in the basis where gluino couplings are diagonal
and quark mass matrices are diagonal. The bounds are particularly strong
on the combination
$$\VEV{\delta^q_{ij}}\equiv\sqrt{(\delta^q_{LL})_{ij}
(\delta^q_{RR})_{ij}}.\eqn\defavdel$$

Using formulae from ref. [\gama] we present the constraints from eqs.
\mixing\ and \radiative\ in Table 1.
\vskip 0.5cm
\centerline{Table 1}
\centerline{Upper bounds on squark mass parameters from FCNC}
\vskip 0.5cm
\begintable
$ \VEV{\delta^d_{12}}$ |$(\delta^d_{MM})_{12}$ |$(\delta^d_{LR})_{12}$ |
$ \VEV{\delta^d_{13}}$ |$(\delta^d_{MM})_{13}$ |$(\delta^d_{LR})_{13}$ |
$(\delta^d_{LR})_{23}$ |
$ \VEV{\delta^u_{12}}$ |$(\delta^u_{MM})_{12}$ |$(\delta^u_{LR})_{12}$  \cr
0.006 | 0.05 | 0.008 | 0.04 | 0.1 | 0.06 | 0.04| 0.04 | 0.1 | 0.06
\endtable

The bounds in Table 1 correspond to $\tilde m=1\ TeV$ and scale like
$\tilde m$. They are given for a gluino mass $m_{\tilde g}=\tilde m$,
and become somewhat weaker (up to a factor of a few) for $m_{\tilde
g}>\tilde m$.  The matrix elements were calculated in the vacuum
insertion approximation and we used $f_B=f_D=0.2\ GeV$ (the bounds on
$(\delta^d_{MN})_{13}$ and $(\delta^u_{MN})_{12}$ scale like $1/f_B$ and
$1/f_D$, respectively).  The hadronic uncertainties (and the dependence
on $\tilde m$ and on ${m_{\tilde g}\over\tilde m}$) imply that the
bounds in Table 1 should be trusted only to within a factor of, say,
3--4.

Bounds on $(\delta^{d}_{LL})_{23}$, $(\delta^{d}_{RR})_{23}$ from
$b\rightarrow s\gamma$ scale like $\tilde m^2$. At $\tilde m\sim1\ TeV$,
the suppression of the squark-gluino electromagnetic penguin diagram due
to the heavy squark mass is strong enough to give no constraint on
$(\delta^{d}_{MM})_{23}$.

The bounds in Table 1 pose a serious problem to generic SUSY models,
where $(\delta^q_{LL})_{ij}$, $(\delta^q_{RR})_{ij}$, are expected to be
of order one, and $(\delta^q_{LR})_{ij}$ is expected to be of order
${m_Z\over\tilde m}$.  The standard solution to this problem is to
assume

$(a)$ Degeneracy (universality): each of the diagonal blocks, $\tilde
M^{q2}_{LL}$ and $\tilde M^{q2}_{RR}$, is proportional to the unit
matrix to a very good approximation;

$(b)$ Proportionality: each of the nondiagonal blocks, $\tilde
M^{q2}_{LR}$, is proportional to the corresponding quark mass matrix
$M^q$ to a very good approximation.

If the conditions of degeneracy and proportionality are fulfilled {\it
exactly}, then the various $\delta^q_{MN}$ are diagonal and there is no
contribution to FCNC processes.  If the two conditions hold at some high
energy scale but are violated by radiative corrections (as assumed in
the minimal supersymmetric Standard Model), then a GIM-like suppression
keeps the contributions within bounds.

These conditions have been known for a while (see {\it e.g.} [\nilles]),
and were discussed by numerous authors, including [\georgi].  In the
context of hidden sector supergravity models they were addressed by
[\weinberg] but in the generic such model, they are unmotivated.  This
is also the case in string theory [\ibanez].  Both degeneracy and
proportionality can be natural if SUSY breaking is communicated to the
light particles by gauge interactions [\dine] or in models with a
nonabelian horizontal symmetry [\dinetoo].

In this work, we would like to suggest an alternative mechanism to
suppress the squark contributions to FCNC: an approximate {\it alignment
of quark and squark mass matrices.} We would first present the mechanism
and then show that it arises naturally in the framework of Abelian
horizontal symmetries.

The idea of quark-squark alignment is rather simple: assume that for
some symmetry reason the matrices $\delta^q$ of eq. \defdelta\ (the
squark mass-square matrices in the basis related by supersymmetry to the
basis in which the quark mass matrices are diagonal) are diagonal.
Then, regardless of whether the squarks are degenerate or not, squark
contributions to FCNC vanish.  In reality, we do not expect such an
exact condition to hold, but it could naturally be a good approximation
that would suppress FCNC to acceptable values.

The implementation of the quark-squark alignment mechanism for the
off-diagonal blocks $\delta^q_{LR}$ is rather simple. If a symmetry
leads to small entries in the quark mass matrices, it will at the same
time lead to small entries in $\tilde M^{q2}_{LR}$.  In addition, a
separation of the SUSY scale $\tilde m$ and the electroweak scale $m_Z$
will give a further suppression factor, ${m_Z\over\tilde m}$. Together,
the two factors typically give $(\delta^q_{LR})_{ij}\sim{\sqrt{m^q_i
m^q_j}\over\tilde m}$, leading to very small contributions to FCNC.

To see explicitly how the approximate alignment works in the diagonal
blocks, we neglect the small left-right mixing due to the nondiagonal
blocks. (This approximation is well-justified in all models discussed
below.) The $\tM^{q2}_{LL}$ and $\tM^{q2}_{RR}$ matrices are
diagonalized by unitary transformations:
$$\eqalign{
\tV^q_L\tM^{q2}_L\tV^{q\dagger}_L=&
{\rm diag}\{\tm^2_{q_{L1}},\tm^2_{q_{L2}},\tm^2_{q_{L3}}\},\cr
\tV^q_R\tM^{q2}_R\tV^{q\dagger}_R=&
{\rm diag}\{\tm^2_{q_{R1}},\tm^2_{q_{R2}},\tm^2_{q_{R3}}\}.\cr
}\eqn\deftv$$
In the basis where mass matrices for both quarks and squarks are
diagonal, gluino interactions depend on mixing matrices
$K_{L,R}^q$:
$$K_L^q=V_L^q\tV_L^{q\dagger},\ \  K_R^q=V_R^q\tV_R^{q\dagger}.
\eqn\twocon$$
The dependence of FCNC processes on squark masses and mixing is
of the form
$$\sum_{\alpha, \beta} (K_M^q)_{i\alpha}
(K_M^q)^*_{j\alpha}(K_N^q)_{i\beta}
(K_N^q)^*_{j\beta}f(\tm^2_{q_{M\alpha}},\tm^2_{q_{N\beta}}),\eqn\delk$$
where $\alpha$ and $\beta$ label the squarks in the loop and
$f(\tm^2_{q_{M\alpha}},\tm^2_{q_{N\beta}})$ is a function of squark
masses.

There are two ways to suppress the off-diagonal terms $(i\neq j)$ in
eq. \delk:

1. The squarks in each sector are degenerate. Then, $f$ is independent
of $\alpha$ and $\beta$ and the sum in eq. \delk\ simplifies,
$$\sum_\alpha(K_M^q)_{i\alpha} (K_M^q)^*_{j\alpha}=0\
{\rm for}\ i\neq j,\eqn\degcon$$
where we used the unitarity of the $K$-matrices (equivalently, in
this case $\tilde V^q$ are arbitrary).  This is the degeneracy
condition discussed above.

2. The matrices $K_M^q$ are close to the unit matrix,
$$(K_M^q)_{ij}\ll1\ {\rm for}\ i\neq j.\eqn\aligncon$$
Notice that for \aligncon\ to be fulfilled, the diagonalizing matrices
for quarks, $V_{L,R}^q$, and the diagonalizing matrices for squarks,
$\tV_{L,R}^q$, have to be approximately equal,
$$V_L^q\tV_L^{q\dagger}\approx{\bf 1},\ \
V_R^q\tV_R^{q\dagger}\approx{\bf 1},\ \ (q=u,d).\eqn\alicon$$
Thus, it should be possible to simultaneously diagonalize (at least
approximately) the quark mass matrices and the squark mass-squared
matrices while preserving diagonal gluino interactions ({\it i.e.} the
diagonalizing matrices act on the quark superfields). This is the
alignment mechanism.

If the squark masses are of the same order of magnitude, $\tm$, but
not degenerate, the expression \delk\ is approximately
$$\left( \max_\alpha(K_M^q)_{i\alpha} (K_M^q)^*_{j\alpha} \right)
\left( \max_\beta (K_N^q)_{i\beta} (K_N^q)^*_{j\beta} \right)
f(\tm^2,\tm^2) . \eqn\delkdet$$
It is easy to see that the matrices $\delta^q_{MM}$ in eq. \defdelta\
are related to the matrices $K$ through
$$(\delta^q_{MM})_{ij} \sim \max_\alpha(K_M^q)_{i\alpha}
(K_M^q)^*_{j\alpha}.  \eqn\delrelk$$
This estimate does not depend on assumptions such as the smallness of
off-diagonal mass terms and therefore also on the smallness of mixing
angles.

The constraints that arise from $K-\bar K$ mixing are the most difficult
to satisfy. They require a particularly precise alignment in the first
two generations of the down sector, namely $(K_M^d)_{12}$ have to be
much smaller than the Cabibbo angle.  This can be ensured if in some
basis $M^d_{1i}$, $M^d_{i1}$ and $(\tM^{d2}_{MM})_{1i}$ for $i=2,3$ are
sufficiently small.  $M^u_{12}$ should then be sufficiently large in the
same basis to produce the Cabibbo mixing.  These conditions guarantee
the alignment in other bases.  In general there is no preferred basis
and therefore such an alignment is not natural.

Quark-squark alignment arises naturally in models based on Abelian
horizontal symmetry.  First, such models have a natural basis for the
quarks in which the alignment is as described in the previous paragraph.
Second, they can naturally explain the hierarchy in the entries and can
thus provide the necessary small numbers. Examples of such models were
first suggested in ref. [\frni].  They have been extensively
investigated in the context of grand unified models in ref. [\dim] and
have recently been studied in the context of low energy supersymmetry in
ref. [\lns].  In this framework the hierarchy in the quark sector
parameters arises from a spontaneously broken horizontal discrete
symmetry.  If the symmetry were exact, only the top quark (and maybe the
bottom quark) would have a mass. However, the symmetry is spontaneously
broken by the VEV $\VEV{S}$ of a Standard Model singlet scalar. This
breaking gives masses and mixings to all other quarks (except, maybe,
the up quark) through nonrenormalizable terms, induced by integrating
out heavy color-triplet fermions of mass $M>\VEV{S}$.  The hierarchy in
mass ratios and mixing angles arises from their dependence on various
powers of the ratio $\e={\VEV{S}\over M}$.

We now give explicit examples of models within this framework with
precise enough alignment.  The models are very similar to the ones
presented in [\lns].  In all of these models, the horizontal symmetry is
a discrete subgroup of
$$H=U(1)_X\times U(1)_{H_1}\times U(1)_{H_2}.\eqn\horsym$$
The $X$ charge of $\phi_d$ is $-1$, that of all $\bar d_i$ is $+1$ and
all other fields have vanishing $X$ charge.  The scalar sector consists,
in addition to the Standard Model doublets $\phi_u$ and $\phi_d$, of two
Standard Model singlets $S_1$ and $S_2$. The ($H_1,H_2$) charges of the
scalars are
$$\phi_u(0,0),\ \ \phi_d(0,0),\ \ S_1(-1,0),\ \ \
S_2(0,-1),\eqn\horcharge$$
with the horizontal symmetry breaking parameters
$$\ea={\VEV{S_1}\over M}\sim0.04,\ \ \
\eb={\VEV{S_2}\over M}\sim0.008.\eqn\horbreak$$
Here $M$ is a scale higher than the scale of symmetry breaking,
$\VEV{S_i}$, which communicates the breaking to the light fermions.  The
small parameters $\ea$ and $\eb$ determine the hierarchy in the quark
sector parameters.  Defining another small parameter $\e\sim0.2$, so
that $\ea\sim\e^2 \sim 0.04 $ and $\eb\sim\e^3 \sim 0.008$, we require
that our models give
$$\eqalign{
\e&\sim|V_{us}|,\cr
\e^2&\sim|V_{cb}|,\ \ {m_d\over m_s},\ \ {m_s\over m_b},\ \
{m_b\over m_t},\cr
\e^3&\sim|V_{ub}|,\ \ {m_u\over m_c},\ \ {m_c\over m_t}.\cr}\eqn\pred$$
(An interesting option is to have $m_u=0$. This solves the strong CP
problem but the phenomenological viability of this option is
controversial. In this paper we choose to study only models that give
$m_u\sim m_d$.)

The horizontal charges of the second and third generation are uniquely
fixed by eq. \pred, but there are two possible $(H_1,H_2)$ assignments
for each of $Q_1$, $\bar u_1$ and $\bar d_1$.  It is on this point that
our models differ from the one explicitly presented in ref. [\lns].
Requiring the alignment discussed above, the horizontal charges $(H_1,
H_2)$ for $Q_1$ and $\bar d_1$ are fixed and we find two examples:
$$\eqalign{ Q_1(3,-1),\ \ Q_2&(1,0),\ \ Q_3(0,0),\cr
\bar u_1(h_1,h_2),\ \ \bar u_2&(-1,1),\ \ \bar u_3(0,0),\cr
\bar d_1(-3,3),\ \ \bar d_2&(1,0),\ \ \bar d_3(1,0),\cr}
\eqn\honehtwoqsa$$
with
$$(h_1,h_2)=\cases{(-3,3)&Model A, \cr (0,1)&Model B.\cr} \eqn\musimmd$$

The horizontal charges \honehtwoqsa\ lead to the following estimates for
the various entries in the quark mass matrices:
$$\eqalign{{M^u({\rm Model\ A})\over\VEV{\phi_u}}\sim
\pmatrix{
\eb^2 & \ea^2 & 0 \cr
0 & \eb & \ea \cr
0 & 0 & 1 \cr},&\ \ \
{M^u({\rm Model\ B})\over\VEV{\phi_u}}\sim
\pmatrix{
\ea^3 & \ea^2 & 0 \cr
\ea\eb & \eb & \ea \cr
\eb & 0 & 1 \cr},
\cr
{M^d\over\VEV{\phi_d}}\sim &\pmatrix{\eb^2 & 0 & 0 \cr
0 & \ea^2 & \ea^2 \cr 0 & \ea & \ea \cr}. \cr}
\eqn\qmqsa$$
The vanishing entries arise because in supersymmetric theories the
Yukawa couplings are analytic in the scalars (we will return to this
below). It is easy to check that the mass matrices \qmqsa\ are
consistent with \pred.  As in the model in [\lns] we explain eight small
dimensionless numbers in terms of two small numbers, $\e_i$, and thus
obtain six order of magnitude mass-angle relations.  However, without
imposing $\ea^3 \sim \eb^2$ in the present model we lose the celebrated
relation $\theta_c\cong \sqrt{m_d \over m_s}$ even as an order of
magnitude one.

An order of magnitude estimate of the various entries in the diagonal
blocks of the squark mass-squared matrices gives
$$\eqalign{
{\tM^2_{LL}\over\tm^2}\sim&\ \pmatrix{1 & \ea^2\eb & \ea^3\eb \cr
\ea^2\eb & 1 & \ea \cr  \ea^3\eb & \ea & 1 \cr},\cr
{\tM^{u2}_{RR}\over\tm^2}\sim&\ \pmatrix{
1 & \ea^{|h_1+1|}\eb^{|h_2-1|} & \ea^{|h_1|}\eb^{|h_2|} \cr
\ea^{|h_1+1|}\eb^{|h_2-1|} & 1 & \ea\eb \cr
\ea^{|h_1|}\eb^{|h_2|} & \ea\eb & 1 \cr},\cr
{\tM^{d2}_{RR}\over\tm^2}\sim&\ \pmatrix{1 & \ea^4\eb^3 & \ea^4\eb^3 \cr
\ea^4\eb^3 & 1 & 1 \cr  \ea^4\eb^3 & 1 & 1 \cr}\cr}
\eqn\sqmqsa$$
(unlike the quarks mass matrices and the nondiagonal blocks, these do
not have to be analytic in the $S$ fields).  For the nondiagonal blocks,
$${(\tM^{q2}_{LR})_{ij}\over\tilde m}\sim (M^q)_{ij}\eqn\ndiagqsa$$
where the $\sim$ sign indicates that the different entries are of
the same order of magnitude.

With the above mass matrices, it is possible to estimate the various
$(\delta_{MN}^q)_{ij}$. As argued above, the $(\delta_{LR}^q)_{ij}$
contributions are very small.  To estimate $(\delta_{MM}^q)_{ij}$ we
note that in our framework all squark masses are of order $\tilde m$ but
not degenerate. Then eq. \delrelk\ holds. Taking into account that all
diagonal elements in the diagonalizing matrices are of order one, we find
$$\eqalign{(\delta_{LL}^d)_{12}\sim{\rm max}\{&
(V_L^d)_{12},\ (\tV^d_L)_{12},\ (V_L^d)_{13}(\tV_L^d)_{23},\cr
\ &(V_L^d)_{23}(\tV_L^d)_{13},\ (V_L^d)_{13}(V_L^d)_{23},\
(\tV_L^d)_{13}(\tV_L^d)_{23}\},\cr}\eqn\estqsa$$
and similarly for the other $(\delta_{MM}^q)_{ij}$.

Our results are summarized in Tables 2 and 3.  We give here the
predictions of the two models discussed above -- model A and model B --
for the various $(\delta^q_{MM})_{ij}$ relevant to FCNC. These
predictions are compared to the phenomenological constraints of Table 1.
As the constraints on $(\delta^q_{LR})_{ij}$ (and on
$(\delta^d_{MM})_{23}$) are fulfilled almost trivially in our framework,
we do not present them in the Tables. Table 2 presents the predictions
for the down sector, which are identical for both models. The
predictions for $D-\bar D$ mixing, presented in Table 3, are different
in the two models, but in both they are close to the experimental
bounds.
\vskip 1cm
\centerline{Table 2}
\centerline{SUSY contributions to down sector FCNC}
\vskip 0.5cm
\begintable
  |
$\VEV{\delta^d_{12}}$ | $(\delta^d_{LL})_{12}$ | $(\delta^d_{RR})_{12}$ |
$\VEV{\delta^d_{13}}$ | $(\delta^d_{LL})_{13}$ | $(\delta^d_{RR})_{13}$
\cr
 Expt. Upper Bound |
$0.006$ | $0.05$ | $0.05$ |
$0.04$ | $0.1$ | $0.1$
\cr
 Models A, B |
 $\ea^3\eb^2$ | $\ea^2\eb$ | $\ea^4\eb^3$ |
$\ea^{7/2}\eb^2$ | $\ea^3\eb$ | $\ea^4\eb^3$
\nr
$\quad$ | $\sim 4\cdot 10^{-9}$ | $\sim 10^{-5}$ | $\sim 10^{-12}$ |
$\sim 10^{-9}$ | $\sim 5 \cdot 10^{-7}$ | $\sim 10^{-12}$
\endtable

\vskip 1cm
\centerline{Table 3}
\centerline{SUSY contributions to up sector FCNC}
\vskip 0.5cm
\begintable
  |
$\VEV{\delta^u_{12}}$ | $(\delta^u_{LL})_{12}$ | $(\delta^u_{RR})_{12}$
\cr
 Expt. Upper Bound | $0.04$ | $0.1$ | $0.1$ \cr
Model A |  $\ea^2/\eb^{1/2}$ | $\ea^2/\eb$ | $\ea^2$ \nr
$\quad$| $\sim 0.02 $ | $\sim 0.2 $ | $\sim 10^{-3}$ \cr
Model B |  $\ea^{3/2}/\eb^{1/2}$ | $\ea^2/\eb$ | $\ea$ \nr
$\quad$| $\sim 0.09 $ | $\sim 0.2 $ | $\sim 0.04$
\endtable

We conclude that these two examples satisfy all the constraints from
FCNC without any degeneracy among squarks.  (Remember that both the
experimental upper bound and the order of magnitude predictions in our
models suffer from ambiguities in multiplicative factors of order one.)

As discussed above, the most difficult condition to satisfy is
$$(V^d_{L})_{12},\ \ (V^d_{R})_{12}\lsim0.006\eqn\require$$
which comes from $K-\bar K$ mixing. It is naturally guaranteed in our
models because the horizontal symmetry forces $M^d_{1i}=0$ and
$M^d_{i1}=0$ for $i=2,3$.  Such vanishing entries arise because of a
combination of the horizontal symmetry and the analyticity of the
superpotential.  Only terms of the form $Q_i\phi_d\bar d_j ({S_1\over
M})^m({S_2\over M})^n$ (with non-negative $n$ and $m$) are
supersymmetric but for these entries such terms are not compatible with
the horizontal symmetry ({\it e.g.} $H_2(Q_1)+ H_2(\bar d_2) <0$).  We
will see below that soft SUSY breaking can generate nonanalytic terms,
but they are suppressed at least by several powers of $\epsilon$'s as
dictated by the horizontal symmetry ({\it e.g.}
$M^d_{12}\sim\ea^4\eb\sim 2\cdot10^{-8}$) which render them harmless.

The entries in the quark mass matrices can be modified by several
mechanisms:

$(i)$ The soft SUSY breaking terms can introduce non-analytic Yukawa
terms with $S^\dagger_i$.  These can arise either at tree level by
integrating out the fields at the scale $M$ and are suppressed by
$m_{3/2}\over M$ or by loops in the low energy theory in which case they
are suppressed by at least $\alpha_s \over \pi$.  As mentioned in the
previous paragraph, the suppression by factors of $\epsilon$'s makes
these very small.

$(ii)$ The full Lagrangian has only a {\it discrete} horizontal
symmetry, say $Z_n$. Then, for non negative $m$ such that $H(Q_1)+H(\bar
d_2)+mH(S_i)=-n$, the term $Q_1\phi_d\bar d_2({S_i\over M})^m$ is
$Z_n$-invariant.  We have to check then that $\e_i^m$ is small enough to
make this contribution harmless.

$(iii)$ The group $U(1)_X$ must be broken because otherwise the model
has an unacceptable axion at the weak scale.  This introduces factors of
$\eta^2={\VEV{\phi_u}\VEV{\phi_d} \over M^2}$ into the mass matrices.
For sufficiently large $M$, the effects of these factors on tree level
FCNC, on the hierarchy and on the alignment are negligible\foot{We thank
M. Leurer for a useful discussion on this point.}.

$(iv)$ In some specific high-energy models, some terms allowed by all
symmetries are not generated by integrating out the massive fields at
tree level.  This can happen if there are too few massive particles to
induce all transitions [\lns].  We have to check that all necessary
terms can indeed be generated.

The high energy spectrum needed in our models consists of at least four
massive fermions of charge +2/3 and five of charge --1/3.  To avoid
Landau poles below the Planck scale, we need either $M\gsim10^7\ TeV$ or
a larger symmetry group at high energies [\lns].

We have constructed explicit models belonging to the class of models
defined by eq. \honehtwoqsa\ but where $U(1)_X$ is broken in the
superpotential by terms of the form $\phi_u\phi_d S_i$ or
$\phi_u\phi_d$ and where only an anomaly-free discrete subgroup is
maintained.  Explicitly, in Model A (B), the anomaly free symmetry is
$Z_n\times Z_5$ with $n\geq3$ ($Z_6\times Z_n$ with $n\geq3$).  In both
models, the zeros in $M^d$ are lifted but the quark-squark alignment
remains accurate enough to satisfy all constraints.

Additional constraints on squark parameters from FCNC arise from
diagrams that involve winos\foot{We thank M. Dine for a useful
discussion on these diagrams.}.  When $SU(2)$ breaking effects in the
squark mass matrices are negligible -- as is the case in our framework
-- the mixing angles for the couplings $\tilde w^- u_i \tilde
d^\dagger_j$ and $\tilde w^+ d_i \tilde u^\dagger_j$ are $(K_L^u)_{ij}$
and $(K_L^d)_{ij}$, respectively. Consequently, for $m_{\tilde w}\sim
m_{\tilde g}$, diagrams with intermediate winos are suppressed by
$\left({g\over g_s}\right)^4$ compared to the LL gluino diagrams.  The
constraints from wino interactions are fulfilled whenever those from
gluino interactions are. (A similar statement can be made on constraints
{}from zino and photino interactions.)

To summarize: models with Abelian horizontal symmetries, originally
constructed to explain the hierarchy in the quark sector parameters, can
naturally align the mass matrices of quarks with the mass-squared
matrices for squarks, to a good approximation. As a result, the
contributions from squark-gluino loop diagrams to neutral meson mixing
and to radiative $B$ decay are suppressed compared to a generic
low-energy supersymmetric framework, even though there is no degeneracy
among squarks.  We would like to stress, however, that our main point is
the alignment.  The models based on horizontal symmetry are only
examples meant to demonstrate that the alignment can be natural.

This class of models where the bounds from $K-\bar K$ mixing are
satisfied have the special feature that the Cabibbo mixing between the
first and second generation arises dominantly from mixing in the up
sector.  (Remember that here, unlike the Standard Model, we have a
natural interaction basis, the one where the horizontal charges are
diagonal.)  This leads to a testable signature of the quark-squark
alignment mechanism, namely that $D-\bar D$ mixing should be close to
the experimental upper bound.  This is in contrast to the Standard
Model, where [\dght] ${\Delta m_D\over m_D}\sim 10^{-15}$, and maybe
even considerably smaller than that if heavy quark symmetry
considerations apply [\georgid].  Of course, the most prominent
prediction (although it will take some time to examine it
experimentally) of our scheme is that, in contradiction with the
conventional wisdom, the squarks are not degenerate.

\vskip 1cm
\centerline{ACKNOWLEDGMENTS}
It is a pleasure to thank T. Banks, A. Dabholkar, M. Dine, K.
Intriligator and M. Leurer for several useful discussions.  We also
thank T. Banks and M. Dine for comments on the manuscript.  YN is
grateful to the theory group at Rutgers University for their
hospitality. YN is an incumbent of the Ruth E. Recu Career Development
Chair and is supported, in part, by the Israel Commission for Basic
Research and by the Minerva Foundation. This work was supported in part
by DOE grant DE-FG05-90ER40559.
\vskip 1cm
\refout
\end